# RF PROCESSING OF X-BAND ACCELERATOR STRUCTURES AT THE NLCTA*

C. Adolphsen, W. Baumgartner, K. Jobe, R. Loewen, D. McCormick,
M. Ross, T. Smith, J.W. Wang, SLAC, Stanford, CA 94309 USA
T. Higo, KEK, Tskuba, Ibaraki, Japan

*Abstract*

During the initial phase of operation, the linacs of the Next Linear Collider (NLC) will contain roughly 5000 X-Band accelerator structures that will accelerate beams of electrons and positrons to 250 GeV. These structures will nominally operate at an unloaded gradient of 72 MV/m. As part of the NLC R&D program, several prototype structures have been built and operated at the Next Linear Collider Test Accelerator (NLCTA) at SLAC. Here, the effect of high gradient operation on the structure performance has been studied. Significant progress was made during the past year after the NLCTA power sources were upgraded to reliably produce the required NLC power levels and beyond. This paper describes the structures, the processing methodology and the observed effects of high gradient operation.

## 1 INTRODUCTION

Over the past four years, four NLC prototype X-Band (11.4 GHz) accelerator structures have been processed to gradients of 50 MV/m and higher at the NLCTA [1]. The structures are traveling wave ($2\pi/3$ phase advance per cell), nearly constant gradient (the group velocity varies from 12% to 3% c) and 1.8 m long (206 cells) with a fill time of 100 ns. They were built in part to test methods of long-range transverse wakefield suppression. In two of the structures (DS1 and DS2), the deflecting modes are detuned, and in the other two (DDS1 and DDS2), they are damped as well [2]. The changes made for these purposes should not affect their performance as high gradient accelerators, which is the focus here. For completeness, results are included from a prototype JLC structure (M2) that was processed at the Accelerator Structure Test Area (ASTA) at SLAC [3]. This detuned structure is shorter (1.3 m, 150 cells) and has a somewhat lower group velocity (10% to 2% c) than the NLC structures.

## 2 FABRICATION AND HANDLING

The cells of four of the five structures where single-diamond turned on a lathe, which produces better than 50 nm rms surface roughness: those of DS1 where turned with poly-crystalline diamond tools which yields a surface roughness of about 200 nm rms. Before assembly, the cells were chemically cleaned and lightly etched in a several step process interleaved with water rinses. For the NLC structures, the steps include degreasing, alkaline soak, acid etch and finally an alcohol bath followed by blow drying with $N_2$. The M2 cells were cleaned in a similar manner but with a weaker acid etch and the final cleaning was done with acetone. The bonding and brazing of the cells into structures were done in a hydrogen furnace for the NLC structures and in a vacuum furnace for M2. Before installation in the NLCTA, the NLC structures were vacuum baked at 450-550 °C for 4-6 days and filled with $N_2$. The M2 structure was filled with $N_2$ after assembly and baked in situ at 250 °C for 2 days after installation in ASTA. Based on the structure geometries and the vacuum pump configuration, the maximum pressure levels in the NLC structures after several days of pumping were estimated to be in the low $10^{-8}$ Torr scale while the M2 pressure was likely in the mid $10^{-8}$ Torr scale.

## 3 RF PROCESSING

Of the five structures, only two (DDS1 and M2) were systematically processed to gradients that were not limited by available power (about 200 MW was needed to produce NLC-like gradients). The other three structures were processed in the NLCTA to gradients of about 50 MV/m and used for beam operation. The bulk of processing was done at 60 Hz with 250 ns pulses (100 ns ramp, 150 ns flat top) at NLCTA and 150 ns square pulses at ASTA. The processing rate was paced by rf breakdown in the structures that reflected power toward the sources (klystrons) and increased the structure vacuum pressures. To protect the klystron windows, the rf power was shut off when more than 5 MW of reflected power was detected just downstream of them. The rf was then kept off for a period of time (30 seconds to many minutes) to allow the structures to pump down. Without this interlock, it is likely that the gas pressure would have built up in the structures over many pulses, causing breakdown on every pulse (the pressure threshold to cause breakdown appears to be in the high $10^{-7}$ Torr scale). Thus, the reflected power limit did not greatly hinder the processing, and in fact it suppressed continuous breakdown (as added protection, the rf was automatically shut off if specific pump pressures were exceeded).

\* Work Supported by DOE Contract DE-AC03-76F00515.

The initial processing of the structures was done manually. That is, the power was slowly increased by an operator who also decided when to reset the power after a trip, for example, by monitoring vacuum pump readings. Such operator oversight was important since breakdown was generally accompanied by large pressure increases in the structures (10 to 100 times higher). Above gradients of about 60 MV/m, however, the pressure increases were typically below a factor of ten and sometimes too small to be detectable. With the smaller pressure increases, automated control of the processing was practical. It was developed for both M2 and DDS1 processing, allowing unattended, around-the-clock operation, which improved the processing efficiency.

For DDS1, the processing control algorithm was typically setup to increase the structure input power by 1% if a reflected power trip did not occur within 2 minutes, and to decrease the power by 2% if a trip occurred within 10 seconds. These times were measured relative to the resetting of the rf power, which was ramped-up over a 30 second period starting immediately after a trip. If a trip occurred between 10 seconds and 2 minutes, the power level was not changed. Also, it would not be increased if any pump pressure readings were above tolerance levels. The algorithm for processing M2 used roughly the same logic.

The automated processing yielded smoothly varying peak power levels when viewed over several hour time intervals. The mean power level depended on the algorithm parameters and on the reflected power trip threshold, which as noted above, was set to protect the klystron windows. When processing DDS1 near its maximum power level, roughly half of the trips were immediately proceeded by one or more consecutive pulses with significant reflected power. The first pulse in the sequence likely initiated a pulse-by-pulse gas build-up that eventually produced reflected power large enough to cause the trip (in many of these cases, the location of the breakdown appears to move upstream in the structure during this sequence of pulses). The distribution of reflected power per pulse was broad and peaked at low values. Above the minimum threshold that could be set, the rate of reflected power pulses was roughly ten times that above the nominal threshold (about 100/hour compared to 10/hour). If the trip threshold were lowered during processing, fewer multi-pulse trips would occur but the increase in the trip rate decreased the steady-state power level. So the nominal threshold was used to expedite processing.

The M2 structure was processed in ASTA during a several week period dedicated for this purpose. An average gradient of 50 MV/m was reached in about 20 hours. Achieving higher gradients, however, took exponentially longer. The processing was stopped after 440 hours at which time a maximum gradient of 85 MV/m had been attained. Prior to its use in NLCTA, DS1 was also processed in ASTA [5]. Again, 50 MV/m was achieved fairly quickly, in about 30 hours. After 200 hours, the maximum source power was reached, which produced a 68 MV/m structure gradient.

In contrast, DDS1 was processed in a piecewise manner over a three year period because of rf power generation and transport limitations (which were eventually overcome), and because it was used for beam operation. Thus, its processing history is harder to quantify. After 55 MV/m was reached using the nominal NLCTA rf pulse (100 ns ramp, 150 ns flat top), the pulse length was shortened (100 ns ramp, 50 ns flat top) in an attempt to speed up processing. Over the course of several hundred hours, the gradient was increased to 73 MV/m, after which it remained nearly unchanged during 300 hours of processing. A 250 ns square pulse was then used to better simulate NLC operation, which immediately reduced the maximum gradient to 70 MV/m. During the last 600 last hours of processing, the gradient has not increased above this level.

## 4 EFFECT OF PROCESSING

Past experience has shown that rf breakdown causes surface damage to the tips of cell irises where the fields are highest (about twice the accelerator gradient). To quantify any changes in the NLC/JLC structures, both visual inspections and rf measurements were made. For the latter, a bead-pull technique was used as the primary means to measure the rf phase profile along the structures relative to the nominal phase advance [4]. Since a nylon string has to be pulled through the structure in this procedure, a noninvasive method of determining the phase profile using beam induced rf was developed. However, the dispersion that occurs during the rf propagation through the structure introduces systematic phase changes, so this technique is best used for measuring relative phase changes.

Table 1 summarizes the processing results from the five structures including the net phase change and the number of cells with discernable phase shifts. The phase changes for all but DDS1 were determined from bead-pull measurement comparisons: the DDS1 value is based on a comparison of the initial bead-pull profile with the latest beam-based measurement. As an example, Figure 1 shows bead-pull measurements of DS2 before and after high power operation. One sees that a phase shift occurs at the upstream end of the structure, which is also true in the other four. In Figure 2, beam-based phase measurements of DDS1 are shown at three different times separated by about 300 hours of processing with the short pulse (the gradient was 60-70 MV/m in the first interval, and 68-73 MV/m in the second). Each curve is the phase profile of the rf induced by the passage of a 20 ns bunch train (11.4 GHz bunch spacing) through the structure. The rf is coupled out at the end of the structure so the rf induced in the first cell comes out last. The large phase increase after 100 ns is due to dispersion. The dispersion also smoothes out any fast phase variations as does the finite length of the bunch train (a 2 ns train is now being used). However, a progressive increase in the phase can be seen. The direction corresponds to an increase in the cell frequencies, which would occur if copper were removed from the tips of the cell irises (removing a 10 μm layer around the curved portion of the irises yields a 1° phase shift per cell in the front half of DDS1 where the irises are 1.0 to 1.5 mm thick). The phase shifts seen in the bead-pull comparisons also correspond to higher frequencies.

Table 1: Structure Processing Summary

| Structure | Hours Operated | Max Grad. (MV/m) | Phase Change (deg.) | # Cells Affected |
|---|---|---|---|---|
| M2 | 440 | 85 | 25 | 70 |
| DS1 | 550 | 54 | 7 | 80 |
| DDS2 | 550 | 54 | 8 | 100 |
| DS2 | 1000 | 50 | 20 | 150 |
| DDS1 | 2700 | 73 | 60 | 120 |

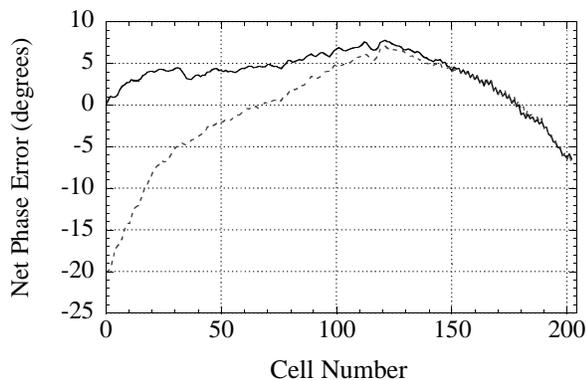 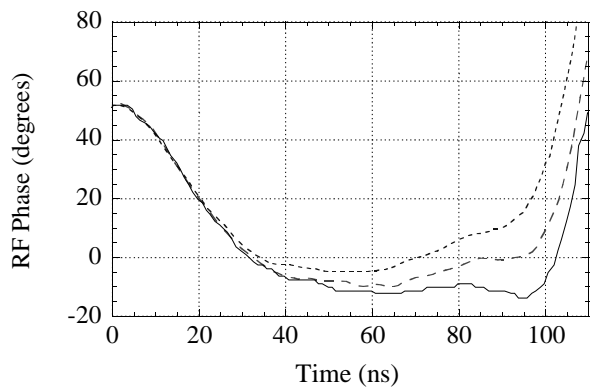

Fig. 1: Bead-pull measurement of the DS2 phase profile before (solid) and after (dotted) 1000 hours of high power operation.

Fig. 2: Phase of beam induced rf measured at different times separated by about 300 hours of processing (first measurement = solid, second = dashed and third = dotted).

In Table 1, one sees that the phase change increases with both gradient and operation time. The DS1 and DDS2 results are interesting to compare, especially since these structures were powered from the same rf source. While DS1 had been processed earlier (see above) and has conventionally machined cells, its phase change during operation in the NLCTA is nearly the same as DS2, which was installed new and has diamond-turned cells. (DS1 was retuned after its processing at ASTA: the phase shift incurred there is not known).

The visual inspections of the structures were done using a boroscope that could access the first and last 30 cells. All structures except DDS1 were examined. Essentially no damage was observed at the downstream end of M2 while the NLC structures showed a small amount of pitting on the tips of the irises in this region. The pits are generally less than 30 μm wide and cover less than a few percent of the surface area. The depths of the pits are hard to estimate, but are probably less than their widths. With this aspect ratio, it takes little rf energy to create them. For example, just $10^{-5}$ of the energy in an rf pulse would vaporize $(30\ \mu m)^3$ of copper if it were converted to heat. For the upstream cells in both M2 and the NLC structures, it looks as if pitting has completely eroded off a layer of the iris surfaces, leaving them covered with 50-100 μm wide dimples. Also, the surface color is a dull silver in contrast to the shiny copper color of the downstream cells.

## 5 CONCLUSION

Five prototype NLC/JLC structures have been processed to high power and show upstream iris damage and phase advance changes at gradients as low as 50 MV/m. One explanation is that the lower rf propagation impedance at the upstream ends of the structures (due to the higher group velocity) leads to more energy being absorbed in breakdown arcs, which act as low impedance loads [6]. This model may also explain why early prototype cavities and structures that had long fill times or low group velocities performed well at high gradients (for example, a 75 cm long NLC structure with 5% group velocity was processed to 90 MV/m without any apparent phase change). A series of low group velocity structures are being built to verify this model as a first step to developing a high gradient version for NLC/JLC. Also, the characteristics of structure breakdown are being extensively studied [7,8].

## 6 REFERENCES

[1]  R. D. Ruth et al., SLAC-PUB-7288 (June 1997).
[2]  J. Wang et al., PAC 99 Proc., p. 3423 (April 1999).
[3]  R. Loewen et al., SLAC-PUB-8399 (June 1997).
[4]  S. Hanna et al., SLAC-PUB-6811 (June 1995).
[5]  J. Wang et al., SLAC-PUB-7243 (August 1996).
[6]  C. Adolphsen, SLAC-PUB-8572 (in progress).
[7]  C. Adolphsen et al., SLAC-PUB-8573 (Sept. 2000), which is an expanded version of this paper.
[8]  Joe Frisch et.al., TUE03, these proceedings.